\documentclass[9pt,twocolumn,twoside]{osajnl}

\journal{optica} 
\usepackage{amssymb}
\newcommand{\gsim}{\raisebox{-0.13cm}{~\shortstack{$>$ \\[-0.07cm]
      $\sim$}}~}

\setboolean{shortarticle}{false}

\title{Laser-induced anti-Stokes fluorescence cooling of ytterbium-doped silica glass by more than 6 Kelvin}

\author[1,2]{Mostafa Peysokhan}
\author[1]{Saeid Rostami}
\author[1,2]{Esmaeil Mobini}
\author[1]{Alexander R. Albrecht}
\author[3]{Stefan Kuhn}
\author[3]{Sigrun Hein}
\author[3]{Christian Hupel}
\author[3]{Johannes Nold}
\author[3]{Nicoletta Haarlammert}
\author[3]{Thomas Schreiber}
\author[3]{Ramona Eberhardt}
\author[4]{Angel S. Flores}
\author[3,5]{Andreas T\"unnermann}
\author[1]{Mansoor Sheik-Bahae}
\author[1,2,*]{Arash Mafi}

\affil[1]{Department of Physics \& Astronomy, University of New Mexico, Albuquerque, New Mexico 87131, USA}
\affil[2]{Center for High Technology Materials, University of New Mexico, Albuquerque, New Mexico 87106, USA}
\affil[3]{Fraunhofer Institute for Applied Optics and Precision Engineering, Albert-Einstein-Str. 7, 07745 Jena, Germany}
\affil[4]{Air Force Research Laboratory, Directed Energy Directorate, 3550 Aberdeen Ave. SE, Kirtland Air Force Base, New Mexico 87117, USA}
\affil[5]{Institute of Applied Physics, Abbe Center of Photonics, Friedrich-Schiller-Universit\"at, Albert-Einstein-Str. 15, 07745 Jena, Germany}
\affil[*]{Corresponding author: mafi@unm.edu}

\begin{abstract}
Laser cooling of a solid is achieved when a coherent laser illuminates the material, and the heat is extracted by resulting anti-Stokes fluorescence. Over the past year, net solid-state laser cooling was successfully demonstrated for the first time in Yb-doped silica glass in both bulk samples and fibers. Here, we improve the previously published results by one order of magnitude and report more than 6K of cooling below the ambient temperature. This result is the lowest temperature achieved in solid-state laser cooling of silica glass to date to the best of our knowledge. We present details on the experiment performed using a 20W laser operating at 1035nm wavelength and temperature measurements using both a thermal camera and the differential luminescence thermometry technique. 
\end{abstract}

\setboolean{displaycopyright}{true}

\begin{document}

\maketitle

\section{Introduction}
The possibility of heat extraction from materials via anti-Stokes fluorescence (ASF) was first suggested by Pringsheim in 1929~\cite{pringsheim1929zwei}. Nearly seven decades later, in 1995, Epstein et al. reported the first experimental observation of laser-induced ASF cooling of a solid in Yb-doped ZBLANP (ZrF\textsubscript{4} –BaF\textsubscript{2} –LaF\textsubscript{3} –AlF\textsubscript{3} –NaF–PbF\textsubscript{2})~\cite{epstein1995observation}. Several attempts have since confirmed laser cooling in various solid-state materials, primarily in rare-earth-doped (RE-doped) crystals and glasses. Laser cooling of RE-doped crystals has been the most successful so far~\cite{seletskiy2010laser, nemova2010laser,seletskiy2016laser}; the record cooling of a Yb-doped YLiF\textsubscript{4} (Yb:YLF) crystal was reported at the University of New Mexico in 2016~\cite{melgaard2016solid}. Several RE-doped glasses have been successfully cooled over the years~\cite{Gosnell:99, PhysRevB.62.3213, hoyt2000observation, thiede2005cooling, fernandez2006anti, nguyen2013towards, peysokhan2019measuring, peysokhan2020characterization}; Yb-doped silica glasses are the most recent additions to the list of successfully cooled RE-doped materials~\cite{mobini2019laser,mobini2020laser, Knall:20, Knall:2020}. Although laser-cooling of RE-doped silica was thought to be elusive over the years, recent investigations pointed out its possibility~\cite{PhysRevApplied.11.014066,10.1117/12.2510889,8426483} and eventually led to its experimental observation~\cite{mobini2019laser,mobini2020laser,10.1117/12.2545233,10.1117/12.2548506, Knall:20, Knall_20_comp, Knall:2020}. In all these reports, the temperature drop of the laser-cooled Yb-doped silica was less than 1\,K. Here we present, to the best of our knowledge, the lowest temperature achieved so far in laser cooling of Yb-doped silica glass by more than 6\,K.

There are many potential applications for optical cooling through ASF. In principle, it can be used for compact, vibration-free refrigeration systems~\cite{Gosnell:99, seletskiy2016laser}, e.g., when precision cooling is demanded in low-thermal-noise detectors and reference cavities of ultra-stable lasers, or even in physiological applications~\cite{zhou2016laser}. One can even envision laser-cooled silica's potential usage as the substrate in silicon photonics devices~\cite{zhu2010high, mobini2019heatcc, jalali2006silicon, soref2010mid}. Another important potential application for ASF cooling is in radiation-balanced fiber lasers (RBFLs), where anti-Stokes fluorescence cooling balances the waste heat generated in the laser~\cite{sbowman, NEMOVA20092571, bowman_min, Nemova:11, Mobini:18, Yang:19}. Historically, RE-doped ZBLAN glasses have been more amenable to the stringent requirements needed for laser cooling. Unfortunately, ZBLAN fibers are low in mechanical durability and chemical stability and hard to cleave and splice, so they are generally less desirable than silica fibers. However, the recent advances in laser cooling of Yb-doped silica glass open a potential pathway for future application in RBFLs. Of course, much more substantial cooling is required to make a viable impact on fiber laser designs; this paper is a step in that direction.  
\section{Review of the recent results}
The cooling efficiency, $\eta_{\rm c}$, characterizes the potential of a material to cool via laser-induced anti-Stokes fluorescence. It is defined as the net power density (per unit volume) extracted from the material ($p_{\rm net}$) per unit total absorbed power density ($p_{\rm abs}$). The cooling efficiency is a function of the pump laser wavelength $\lambda_{\rm p}$ and can be expressed as~\cite{mobini2020laser}
\begin{align}
\label{Eq:cooleff}
\eta_{\rm c}(\lambda_{\rm p})=\dfrac{p_{\rm net}}{p_{\rm abs}}=\dfrac{\lambda_{\rm p}}{\lambda_{\rm f}}\eta_{\rm ext}\,\eta_{\rm abs}(\lambda_{\rm p})-1.
\end{align}
The mean fluorescence wavelength is represented as $\lambda_{\rm f}$. The external quantum efficiency, $\eta_{\rm ext}$, and the absorption efficiency, $\eta_{\rm abs}$, are defined as:  
\begin{align}
\eta_{\rm ext}=\dfrac{\eta_{\rm e} W_{\rm r}}{\eta_{\rm e} W_{\rm r}+W_{\rm nr}},\qquad 
\eta_{\rm abs}(\lambda_{\rm p})=\dfrac{\alpha_{\rm r}(\lambda_{\rm p})}{\alpha_{\rm r}(\lambda_{\rm p})+\alpha_{\rm b}},
\end{align}
where $W_{\rm r}$ and $W_{\rm nr}$ are radiative and non-radiative decay rates of the excited state in RE dopant, respectively, and $\eta_{\rm e}$ is the fluorescence extraction efficiency. $\alpha_{\rm b}$ is the background absorption coefficient and $\alpha_{\rm r}(\lambda_{\rm p})$ is the resonant absorption coefficient due to the RE dopants. Note that the attenuation due to scattering, including Rayleigh scattering, does not contribute to the material's heating; therefore, $\alpha_{\rm b}$ represents only the background absorption and not the total parasitic attenuation.    

For net solid-state optical refrigeration, the cooling efficiency must be positive; therefore, we must show that $\eta_{\rm c}>0$ is reachable over a range of $\lambda_{\rm p}$. The laser pump wavelength $\lambda_{\rm p}$ cannot be much longer than $\lambda_{\rm f}$; otherwise, the pump absorption cross-section would become too small. This would result in a small $\alpha_{\rm r}$ and hence a small $\eta_{\rm abs}$ and a negative cooling efficiency. 
In practice, to observe net cooling, $\lambda_{\rm p}$ can only be slightly longer than $\lambda_{\rm f}$, and both $\eta_{\rm ext}$ and $\eta_{\rm abs}$ must be near unity. To realize the $\eta_{\rm abs}\sim 1$ limit for $\lambda_{\rm p}\gsim\lambda_{\rm f}$, one must increase the RE dopant density to achieve $\alpha_{\rm r}(\lambda_{\rm p})\gg\alpha_{\rm b}$. However, increasing the RE dopant density results in an increase in the non-radiative decay rate, $W_{\rm nr}$, primarily because of the RE clustering and quenching, hence decreasing the external quantum efficiency, $\eta_{\rm ext}$. This unfortunate circle of undesirable influences was recently overcome in Yb-doped silica--it was shown that by adding certain modifiers such as Al, P, F, and Ce, the quenching concentration of silica glass could be increased significantly~\cite{Jesper,Arai,PhysRevApplied.11.014066}. The result was the successful cooling of high-Yb-concentration silica as a fiber preform by Mobini et al.~\cite{mobini2019laser,mobini2020laser} up to 0.7\,K, and as an optical fiber by Knall et al.~\cite{Knall:20} up to 50\,mK.

\begin{figure}[htbp]
\centering
    \includegraphics[width=1.0\linewidth]{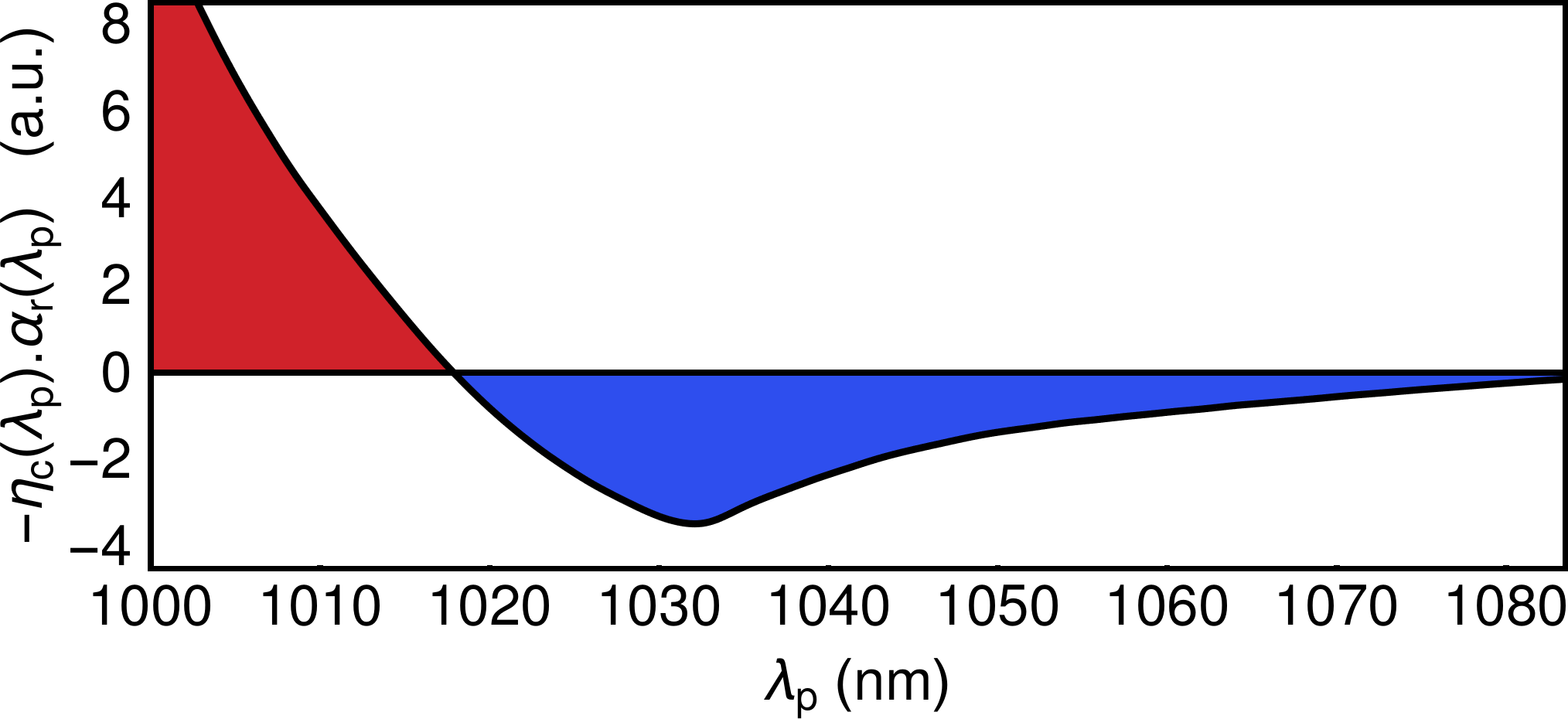}
\caption{This figure shows the value of $-\eta_c\,\alpha_r(\lambda)$, which is proportional to $\Delta T$ at a fixed input laser power (in the low-absorption regime), versus the pump laser wavelength for our sample.}
\label{Fig:OptDeltaT}
\end{figure}
To investigate the cooling efficiency as a function of the pump wavelength and obtain the optimum value of $\lambda_{\rm p}$ for maximum cooling, we performed laser-induced thermal modulation spectroscopy (LITMoS) test~\cite{seletskiy2016laser,Rostami:19} on our Yb-doped silica samples~\cite{mobini2020laser}. In Fig.~\ref{Fig:OptDeltaT}, we show $-\eta_{\rm c} \alpha_{\rm r}$ as a function of $\lambda_{\rm p}$ for the sample used in this paper (same as sample A in Ref.~\cite{mobini2020laser} but some of the cladding is removed; see Table~\ref{tab:values}). This quantity is proportional to the change in the sample temperature for a fixed pump laser power. Figure~\ref{Fig:OptDeltaT} shows that the maximum temperature drop can be obtained at around 1035\,nm. At the time when we carried out our experiments for Refs.~\cite{mobini2019laser,mobini2020laser}, the only viable high-power source in our laboratory was a 1053\,nm laser. In this paper, as we will explain later, we are using a $\lambda_{\rm p}=1035$\,nm source to achieve a higher temperature drop. 
\section{Power cooling experiment}
The samples that we laser-cooled in our experiments reported in Refs.~\cite{mobini2019laser,mobini2020laser} were surrounded by undoped (no Yb-doping) silica glass cladding regions, which provides significant thermal load. The cooling in our experiments was achieved in spite of this large thermal load. For this work, we chose sample A used in Ref.~\cite{mobini2020laser} and removed most of its undoped cladding region to reduce the thermal load and enhance the cooling effect. Moreover, we built a high-power source at the optimum cooling wavelength of 1035\,nm as described below. 

The characteristics of the (fiber preform) sample are listed in Table~\ref{tab:values}. The Yb$_2$O$_3$ concentration is measured by electron probe micro-analysis. The Yb density is calculated from the measured Yb$_2$O$_3$ concentration. The error for the Yb$_2$O$_3$ concentration is related to the applied method's uncertainty in this concentration range. OH$^-$ concentration and parasitic background absorption ($\alpha_b$) are measured by the cut-back method in the fiber form, for which the errors express the repeatability of the measurement setup.
\begin{table}
\centering
\caption{\bf Properties of the Yb-doped silica glass sample}
\begin{tabular}{ccc}
\hline
Parameter & Value & Error\\
\hline
Codopants & Al, P & -\\

Yb$_2$O$_3$ [mol\%] & 0.12 & $\pm 0.01$\\
Yb density [10$^{25}$ m$^{-3}$] & 5.3 & $\pm 0.4$\\
OH$^-$ concentration [ppm] & 3.0 & $\pm 0.5$\\
Core diameter [mm] & 1.7 & $\pm 0.1$\\
Cladding diameter [mm] & 2.9 & $\pm 0.1$\\
Length [mm] & 15.1 & $\pm 0.1$\\
$\alpha_b$ [dB km$^{-1}$] & 10 & $\pm 2$
\end{tabular}
  \label{tab:values}
\end{table} 

\begin{figure*}[ht]
\centering
\includegraphics[width=1.0\linewidth]{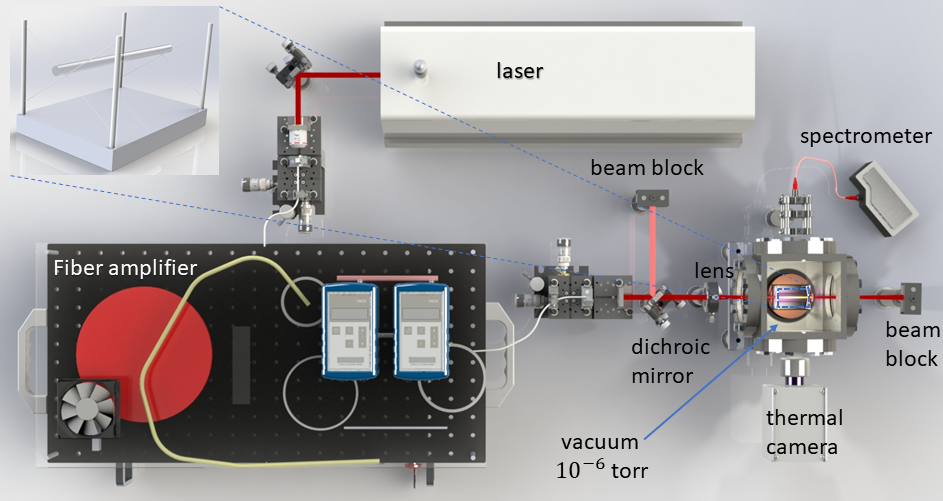}
\caption{A wavelength-tunable continuous-wave Ti:Sapphire laser is coupled to a homemade fiber amplifier's input through a 20X microscope objective. The amplified laser light is collimated again by a lens with the focal length of f=5\,cm. The collimated light is then filtered by a stack of two one-micron long-pass dichroic mirrors. The filtered and collimated light is coupled to the Yb-doped silica glass sample by a lens with the focal length of f=12\,cm. The sample is held inside a vacuum chamber. The upper-left inset shows a sketch of the Yb-doped silica glass sample supported by a set of thin silica fibers to minimize the heat load.}
\label{fig:setup}
\end{figure*}
To make a high-power source at the nearly optimum $\lambda_{\rm p}=1035$\,nm wavelength, we have designed and built a fiber amplifier to amplify the output of our continuous-wave tunable Ti:Sapphire laser. The fiber amplifier's gain medium is a 1.2\,m piece of Yb-doped double-cladding fiber pumped by a high-power diode laser at the wavelength of 976\,nm. The amplifier's input is approximately 300\,mW, and the amplified output of the fiber amplifier is on the order of $\sim$20\,W at 1035\,nm wavelength. We note that any residual diode pump power at the 976\,nm wavelength can be a significant source of heating in the material because the Yb-silica sample's absorption peaks at 976\,nm. Therefore to observe laser cooling, a spectrally pure laser light is essential. To reduce the fiber amplifier's 976\,nm pump leakage in the output as much as possible, we implement a cladding mode stripper scheme at the fiber amplifier's end. We also use a stack of two 1000\,nm wavelength long-pass dichroic mirrors to filter out the rest of the 976\,nm pump leakage.

The experimental setup for the power cooling experiment is shown in Fig.~\ref{fig:setup}. The Ti:Sapphire laser is tuned to 1035\,nm wavelength, which is then amplified by the fiber amplifier. The output laser light is then collimated and filtered. The collimated light is coupled to the sample through a long-focal-length lens from the outside of the vacuum chamber. The vacuum chamber pressure is maintained at $10^{-6}$ torr during the power cooling experiment to minimize the convective heat transfer. A spectrometer captures the sample's fluorescence through a KCl salt window mounted in the chamber. Likewise, the thermal images are recorded via a thermal camera through the thermally transparent KCl salt window, and the images are post-processed to measure the changes in the sample's temperature~\cite{mobini2020laser}. The mean fluorescence wavelength is calculated from the fluorescence emission measured by an optical spectrum analyzer and is found to be $\lambda_{\rm f}=1010$\,nm. In order to decrease the thermal contact on the sample, the sample is mounted on very thin glass fibers. To minimize the back-reflection into the fiber amplifier, the sample and chamber windows are tilted slightly, and a beam block is used to capture the laser after exiting the sample.

\begin{figure}[h]
\centering
    \includegraphics[width=1.0\linewidth]{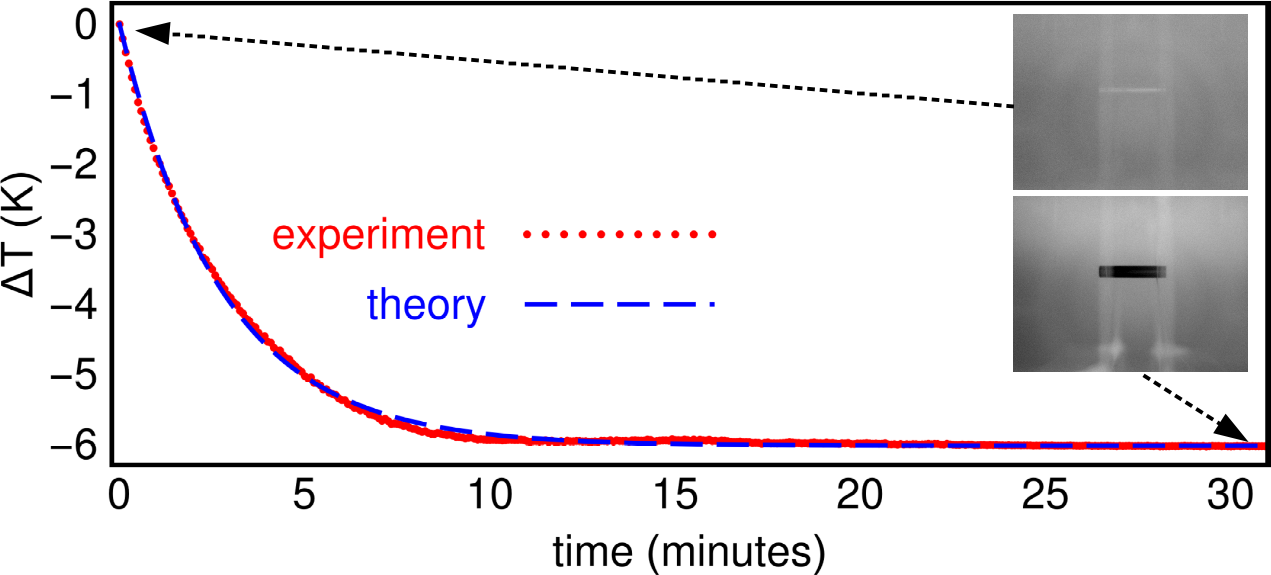}
\caption{The sample's temperature change is plotted as a function of time when exposed to the high-power 1035\,nm laser light. The red dots correspond to the experimental results, and the blue dashed line represents the fitting of the exponential function in Eq.~(\ref{Eq:TemporalTemDffeq-Sol}) to the experimental data. The insets show two thermal images corresponding to before laser exposure and after the final temperature stabilization, respectively.}
\label{Fig:tempvstime}
\end{figure}
The red dots in Fig.~\ref{Fig:tempvstime} show the sample's temperature evolution, measured by the thermal camera, as a function of exposure time to the 20\,W laser light at 1035\,nm. The temperature drop is $\Delta T=T_{s}-T_{0}$, where $T_{s}$ is the sample temperature and $T_0 \approx 23$\,C\textdegree{} is the ambient temperature. The thermal camera saturates at $\Delta T \approx -6\,{\rm K}$, so the temperature may have dropped below the saturation value (see section~\ref{sec:DLT}). In this power cooling experiment, the sample's temperature evolution as a function of time follows the following exponential form (see Ref.~\cite{mobini2020laser} for a derivation):
\begin{align}
\label{Eq:TemporalTemDffeq-Sol}
\Delta T(t)=\Delta T_{\rm max}(e^{-t/\tau_{\rm c}}-1),
\end{align}
where we use the following definitions:
\begin{align}
\label{Eq:DTmax}
\Delta T_{\rm max}=\eta_{\rm c}\dfrac{P_{\rm abs}}{4\epsilon \sigma T_{0}^3 A}, \quad\quad \tau_{\rm c}=\dfrac{\rho V c_{\rm v}}{4\epsilon \sigma T_{0}^3 A}.
\end{align}
$V$ is the sample volume, $\epsilon=0.85$ is the emissivity of the implemented Yb-doped silica glass fiber preform, $\sigma=5.67\times 10^{-8}\,{\rm W}\cdot{\rm m}^{-2}\cdot{\rm K}^{-4}$ is the Stefan-Boltzmann constant,
$T_{0}$ is the ambient temperature, 
$A$ is the surface area of the sample, 
$\rho=2.2 \times 10^{3}\,{\rm kg}\cdot{\rm m}^{-3}$ is the silica glass mass density, 
and $c_{v}=741\,{\rm J}\cdot{\rm kg}^{-1}\cdot {\rm K}^{-1}$ is the specific heat of the silica glass.~\cite{yoder,karimi2018theoretical,mafi:20}. $P_{\rm abs}$ is the absorbed laser power that can be estimated from the Beer-Lambert law in a single pass~\cite{powerl1998physics,Peysokhan:19}:
\begin{align}
P_{\rm abs}=P_{\rm in}\left(1-e^{-\alpha_{\rm r}(\lambda_{\rm p}) l}\right)\approx P_{\rm in}\alpha_{\rm r}(\lambda_{\rm p}) l.
\label{Eq:Pabs}
\end{align} 
$P_{\rm in}$ is the input power coupled into the fiber preform at $z=0$ and $l$ is the sample length. In fact, by combining Eqs.~(\ref{Eq:DTmax}) and~(\ref{Eq:Pabs}), we can see that $\Delta T_{\rm max}\propto \eta_{\rm c}\alpha_{\rm r}$, which is the vertical axis in Fig.~\ref{Fig:OptDeltaT} used to estimate the optimum pump laser wavelength. By fitting the exponential form in Eq.~(\ref{Eq:TemporalTemDffeq-Sol}) to the experimental data (red dots) in Fig.~\ref{Fig:tempvstime}, we obtain $\Delta T_{\rm max}=6.02\pm 0.01\,{\rm K}$ and $\tau_{\rm c}\approx 166\pm 1 \,{\rm s}$--the error-bars are estimated by the fitting procedure. The dashed blue line is the theoretical fit and agrees with the experiment quite well. Using the measured value of $\eta_c\approx 0.016$ at $\lambda_p\approx 1035\,{\rm nm}$ reported in Ref.~\cite{mobini2020laser} for Sample A, we use Eqs.~(\ref{Eq:DTmax}) and~(\ref{Eq:Pabs}) to estimate $\Delta T_{\rm max}\approx 9\,{\rm K}$. This theoretical estimate is consistent with the measured fitted value of $\Delta T_{\rm max}=6.02\pm 0.01\,{\rm K}$, because the heat conduction from the fiber-holder contact and also the parasitic heating from fiber facet imperfections are not included in the theoretically ideal form of Eq.~(\ref{Eq:DTmax}). Moreover, the fitted value for $\tau_{\rm c}$ agrees quite well with the measurement reported in Ref.~\cite{mobini2020laser}, once the difference in geometry is taken into account ($\tau_{\rm c}\approx 175\,{\rm s}$ versus $\tau_{\rm c}\approx 166\,{\rm s}$). The goodness of the fitting in Fig.~\ref{Fig:tempvstime} indicates that despite the satuartion of the camera, the actual value of $\Delta T_{\rm max}$ cannot be much larger than $6\,{\rm K}$.   
\section{Differential Luminescence Thermometry}
\label{sec:DLT}
The thermal camera that we use in our experiments gets saturated at around 6\,K below the ambient temperature and the pixels become black. To cross-check the temperature measurements and see if the sample temperature drops more than 6\,K below the ambient temperature, we use an alternative temperature measurement method dubbed as Differential Luminescence Thermometry (DLT). In this technique, the variation in luminescence intensity distribution with temperature is used to determine the sample's temperature. This variation is due to the temperature-dependence of the Boltzmann population of the crystal field levels of the emitting state and the homogeneous linewidth of the individual crystal field transitions~\cite{Patterson:10}. DLT has been successfully used to measure temperature variations on the order of tens of Kelvin~\cite{melgaard2016solid}; however, it can be quite noisy and less accurate for smaller temperature variations such as those reported here. The reason is that unlike semiconductors where substantial spectral shifts are observed as a function of the temperature~\cite{Babakdlt}, the 4f electrons in REs are shielded from the environment in a solid.

For DLT, the temperature-dependent emission spectral density $S(\lambda,T)$ is obtained in real-time and is referenced to a spectrum at the starting temperature $T_0$. The normalized differential spectrum is defined as:
\begin{align}
    \Delta S(\lambda,T,T_0)= \frac{S(\lambda,T)}{S_{\rm max}(T)} - \frac{S(\lambda,T_0)}{S_{\rm max}(T_0)}.
\end{align}
Normalization to the spectral peak $S_{\rm max}$ is performed to eliminate the effect of input power fluctuations. The scalar DLT signal is given by
\begin{align}
{\rm S}_{\rm DLT}(T,T_0)=\int_{\lambda_1}^{\lambda_2} d\lambda \lvert \Delta S(\lambda,T,T_0) \lvert,
\end{align}
where the limits of integration bracket the sample's spectral emission, eliminating possible contributions from the spurious laser line scattering; we choose $\lambda_1=895\,{\rm nm}$ and $\lambda_2=955\,{\rm nm}$. The temperature drop from the ambient, $\Delta T$, is linearly proportional to ${\rm S}_{\rm DLT}$: $\Delta T = \gamma\cdot{\rm S}_{\rm DLT}$, where $\gamma$ is the proportionality constant. To use DLT for temperature measurements, we first perform a calibration measurement by mounting the sample on a variable-temperature cold plate while pumping the sample with the Ti:Sapphire laser and collecting the spectrum. We find that for our sample $\gamma=-34\pm 2\,{\rm K}$.

\begin{figure}[!h]
\centering
\includegraphics[width=1.0\linewidth]{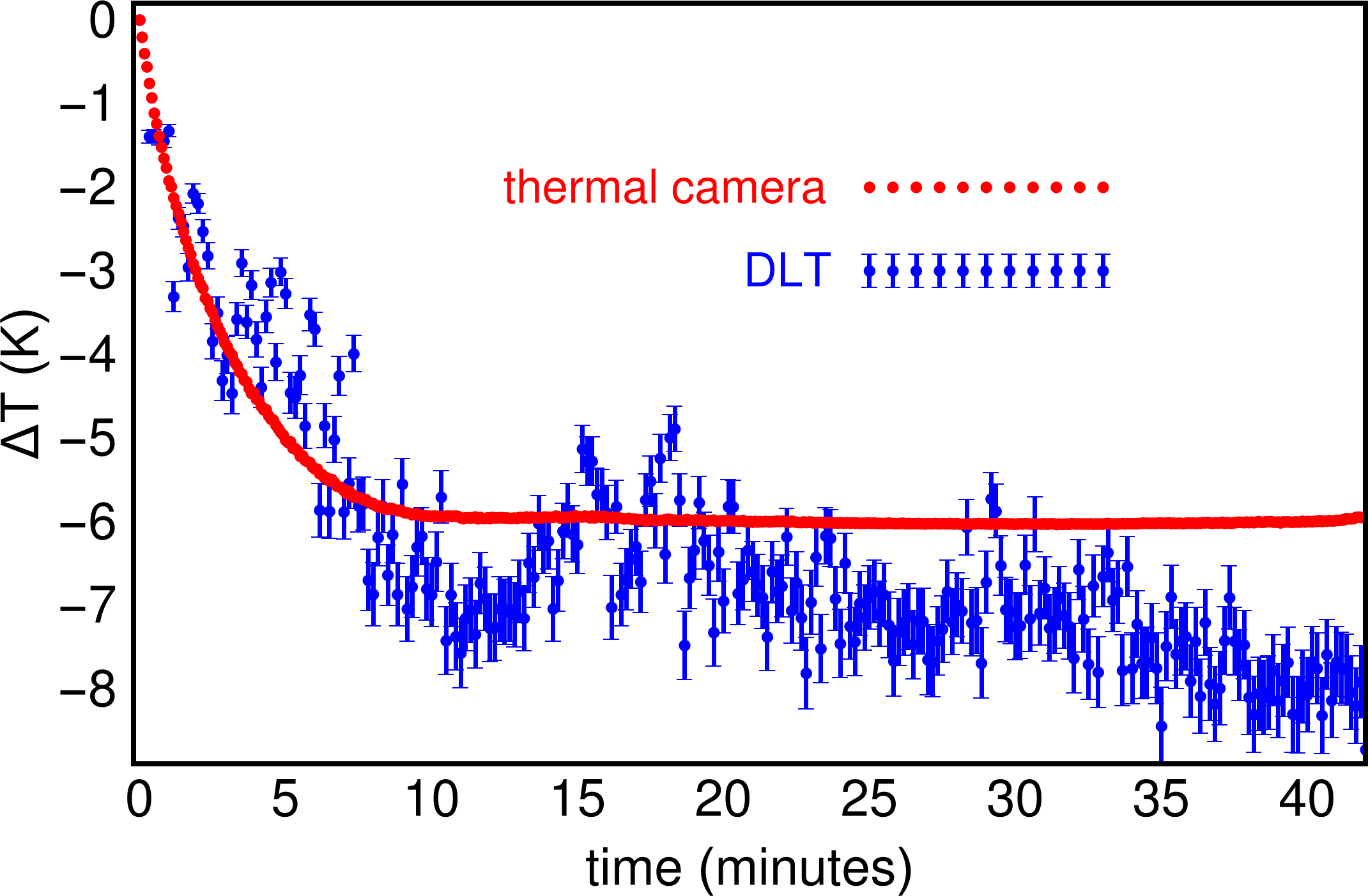}

\caption{The sample's temperature change is plotted as a function of time when exposed to the high-power 1035\,nm laser light. The bue dots are based on the DLT method and the red dots represent the temperature measurements using the thermal camera.}
\label{fig:TempVsTimeDLT}
\end{figure}

We use the DLT calibration result to measure the sample's temperature evolution over time while being exposed to the 20\,W laser light at 1035\,nm by collecting the emission spectral density every ten seconds. The results are shown in Fig.~\ref{fig:TempVsTimeDLT}. The DLT data points are in blue dots where the error-bars are due to the error in $\gamma$ as estimated from the calibration. The results are compared with the thermal camera measurements in red dots. The DLT results are quite noisy as expected; however, the trend agrees with the temperature values from the thermal camera and also hint that the sample is cooled slightly more than 6\,K below the ambient temperature, consistent with the results presented in the previous section. 
\section{Conclusions}
Here we present, to the best of our knowledge, the lowest temperature achieved so far in laser cooling of Yb-doped silica glass by more than 6\,K. This result constitutes almost an order of magnitude improvement compared with the previous result of 0.7\,K reported by  Mobini et al.~\cite{mobini2019laser,mobini2020laser}. The improvement was achieved by using a laser light at the near-optimal wavelength of 1035\,nm (rather than 1053\,nm), nearly doubling the pump power to 20\,W, and reducing the undoped cladding diameter of the fiber preform from 10.7\,mm to 2.9\,mm hence decreasing the thermal load. Future improvements are possible by increasing the pump power and implementing a multi-pass scheme, and improving material specifications. A video clip of the cooling evolution of the sample is presented as the Supplementary. The video shows the temporal evolution of the sample's temperature as captured by the thermal camera in the high-power laser cooling experiment. The thermal image of the sample gets darker as the sample cools due to the exposure to high-power laser. 

\section*{Funding}
This material is based upon work supported by
the Air Force Office of Scientific Research under award number
FA9550-16-1-0362 titled Multidisciplinary Approaches to
Radiation Balanced Lasers (MARBLE).
\section*{Acknowledgments}
The authors would like to acknowledge R. I. Epstein, M. P. Hehlen, and S. D. Melgaard for helpful discussions. 
\section*{Author contributions}
M.P. and A.M. wrote the manuscript and all authors contributed to its final editing. M.P., A.A, and S.R. conducted all the experiments and analyzed the data; S.K., S.H., C.H., J.N., N.H., T.S., and R.E. are responsible for the production and characterization of the silica glass preforms and A.T. supervised their work. A.F. and E.M. helped in making the fiber amplifier. A.M. and M.S.-B. led and supervised the laser cooling aspects of the work and participated in the data analysis.\\

\noindent\textbf{Disclosures.} The authors declare no conflicts of interest.


\begin{thebibliography}{10}
\newcommand{\enquote}[1]{``#1''}

\bibitem{pringsheim1929zwei}
P.~Pringsheim, \enquote{Zwei bemerkungen {\"u}ber den unterschied von
  lumineszenz-und temperaturstrahlung,} {\protect\JournalTitle{Z. Phys.}} \textbf{57}, 739--746 (1929).

\bibitem{epstein1995observation}
R.~I. Epstein, M.~I. Buchwald, B.~C. Edwards, T.~R. Gosnell, and C.~E. Mungan,
  \enquote{Observation of laser-induced fluorescent cooling of a solid,}
  {\protect\JournalTitle{Nature}} \textbf{377}, 500--503 (1995).

\bibitem{seletskiy2010laser}
D.~V. Seletskiy, S.~D. Melgaard, S.~Bigotta, A.~Di~Lieto, M.~Tonelli, and
  M.~Sheik-Bahae, \enquote{Laser cooling of solids to cryogenic temperatures,}
  {\protect\JournalTitle{Nat. Photonics}} \textbf{4}, 161--164 (2010).

\bibitem{nemova2010laser}
G.~Nemova and R.~Kashyap, \enquote{Laser cooling of solids,}
  {\protect\JournalTitle{Rep. Prog. Phys.}} \textbf{73}, 086501
  (2010).

\bibitem{seletskiy2016laser}
D.~V. Seletskiy, R.~Epstein, and M.~Sheik-Bahae, \enquote{Laser cooling in
  solids: advances and prospects,} {\protect\JournalTitle{Rep. Prog. Phys.}} \textbf{79}, 096401 (2016).

\bibitem{melgaard2016solid}
S.~D. Melgaard, A.~R. Albrecht, M.~P. Hehlen, and M.~Sheik-Bahae,
  \enquote{Solid-state optical refrigeration to sub-100 Kelvin regime,}
  {\protect\JournalTitle{Sci. Rep.}} \textbf{6}, 20380 (2016).

\bibitem{Gosnell:99}
T.~R. Gosnell, \enquote{Laser cooling of a solid by 65 k starting from room
  temperature,} {\protect\JournalTitle{Opt. Lett.}} \textbf{24}, 1041--1043
  (1999).

\bibitem{PhysRevB.62.3213}
J.~Fern\'andez, A.~Mendioroz, A.~J. Garc\'{\i}a, R.~Balda, and J.~L. Adam,
  \enquote{Anti-Stokes laser-induced internal cooling of
  ${\mathrm{yb}}^{3+}$-doped glasses,} {\protect\JournalTitle{Phys. Rev. B}}
  \textbf{62}, 3213--3217 (2000).

\bibitem{hoyt2000observation}
C.~Hoyt, M.~Sheik-Bahae, R.~Epstein, B.~Edwards, and J.~Anderson,
  \enquote{Observation of anti-Stokes fluorescence cooling in thulium-doped
  glass,} {\protect\JournalTitle{Physical Review Letters}} \textbf{85}, 3600
  (2000).

\bibitem{thiede2005cooling}
J.~Thiede, J.~Distel, S.~Greenfield, and R.~Epstein, \enquote{Cooling to 208 K
  by optical refrigeration,} {\protect\JournalTitle{Applied Physics Letters}}
  \textbf{86}, 154107 (2005).

\bibitem{fernandez2006anti}
J.~Fernandez, A.~J. Garcia-Adeva, and R.~Balda, \enquote{Anti-Stokes laser
  cooling in bulk erbium-doped materials,} {\protect\JournalTitle{Physical
  review letters}} \textbf{97}, 033001 (2006).

\bibitem{nguyen2013towards}
D.~T. Nguyen, R.~Thapa, D.~Rhonehouse, J.~Zong, A.~Miller, G.~Hardesty, N.-H.
  Kwong, R.~Binder, and A.~Chavez-Pirson, \enquote{Towards all-fiber optical
  coolers using Tm-doped glass fibers,} in \emph{Laser Refrigeration of Solids
  VI,}  vol. 8638 (International Society for Optics and Photonics, 2013), p.
  86380G.

\bibitem{peysokhan2019measuring}
M.~Peysokhan, E.~Mobini, and A.~Mafi, \enquote{Measuring the anti-Stokes
  cooling parameters of a Bb-doped ZBLAN fiber for radiation balancing,} in
  \emph{Sixth International Workshop on Specialty Optical Fibers and Their
  Applications (WSOF 2019),}  vol. 11206 (2019), pp. 112061Q--1.

\bibitem{peysokhan2020characterization}
M.~Peysokhan, E.~Mobini, A.~Allahverdi, B.~Abaie, and A.~Mafi,
  \enquote{Characterization of Yb-doped ZBLAN fiber as a platform for
  radiation-balanced lasers,} {\protect\JournalTitle{Photonics Research}}
  \textbf{8}, 202--210 (2020).

\bibitem{mobini2019laser}
E.~Mobini, S.~Rostami, M.~Peysokhan, A.~Albrecht, S.~Kuhn, S.~Hein, C.~Hupel,
  J.~Nold, N.~Haarlammert, T.~Schreiber \emph{et~al.}, \enquote{Laser cooling
  of silica glass,} {\protect\JournalTitle{arXiv preprint arXiv:1910.10609}}
  (2019).

\bibitem{mobini2020laser}
E.~Mobini, S.~Rostami, M.~Peysokhan, A.~Albrecht, S.~Kuhn, S.~Hein, C.~Hupel,
  J.~Nold, N.~Haarlammert, T.~Schreiber \emph{et~al.}, \enquote{Laser cooling
  of ytterbium-doped silica glass,} {\protect\JournalTitle{Commun. Phys.}} \textbf{3}, 134 (2020).

\bibitem{Knall:20}
J.~Knall, P.-B. Vigneron, M.~Engholm, P.~D. Dragic, N.~Yu, J.~Ballato,
  M.~Bernier, and M.~J.~F. Digonnet, \enquote{Laser cooling in a silica optical
  fiber at atmospheric pressure,} {\protect\JournalTitle{Opt. Lett.}}
  \textbf{45}, 1092--1095 (2020).

\bibitem{Knall:2020}
J.~Knall, M.~Engholm, J.~Ballato, P.~D. Dragic, N.~Yu, and M.~J.~F. Digonnet,
  \enquote{Experimental comparison of silica fibers for laser cooling,}
  {\protect\JournalTitle{Opt. Lett.}} \textbf{45}, 4020--4023 (2020).

\bibitem{PhysRevApplied.11.014066}
E.~Mobini, M.~Peysokhan, B.~Abaie, M.~P. Hehlen, and A.~Mafi,
  \enquote{Spectroscopic investigation of Yb-doped silica glass for
  solid-state optical refrigeration,} {\protect\JournalTitle{Phys. Rev.
  Appl.}} \textbf{11}, 014066 (2019).

\bibitem{10.1117/12.2510889}
J.~M. Knall, A.~Arora, P.~D. Dragic, J.~Ballato, M.~Cavillon, T.~Hawkins,
  S.~Jiang, T.~Luo, M.~Bernier, and M.~Digonnet, \enquote{{Experimental
  investigations of spectroscopy and anti-Stokes fluorescence cooling in
  Yb-doped silicate fibers},} in \emph{Photonic Heat Engines: Science and
  Applications,}  vol. 10936 D.~V. Seletskiy, R.~I. Epstein, and
  M.~Sheik-Bahae, eds., International Society for Optics and Photonics (SPIE,
  2019), pp. 40 -- 49.

\bibitem{8426483}
E.~{Mobini}, M.~{Peysokhan}, B.~{Abaie}, and A.~{Mafi}, \enquote{Investigation
  of solid state laser cooling in ytterbium-doped silica fibers,} in \emph{2018
  Conference on Lasers and Electro-Optics (CLEO),}  (2018), pp. 1--2.

\bibitem{10.1117/12.2545233}
E.~Mobini, S.~Rostami, M.~Peysokhan, A.~R. Albrecht, S.~Kuhn, S.~Hein,
  C.~Hupel, J.~Nold, N.~Haarlammert, T.~Schreiber, R.~Eberhardt,
  A.~T\"unnermann, M.~Sheik-Bahae, and A.~Mafi, \enquote{{Observation of
  anti-Stokes fluorescence cooling of ytterbium-doped silica glass (Conference
  Presentation)},} in \emph{Photonic Heat Engines: Science and Applications
  II,}  vol. 11298 D.~V. Seletskiy, R.~I. Epstein, and M.~Sheik-Bahae, eds.,
  International Society for Optics and Photonics (SPIE, 2020).

\bibitem{10.1117/12.2548506}
J.~M. Knall, P.-B. Vigneron, M.~Engholm, P.~D. Dragic, N.~Yu, J.~Ballato,
  M.~Bernier, and M.~Digonnet, \enquote{{Experimental observation of cooling in
  Yb-doped silica fibers},} in \emph{Photonic Heat Engines: Science and
  Applications II,}  vol. 11298 D.~V. Seletskiy, R.~I. Epstein, and
  M.~Sheik-Bahae, eds., International Society for Optics and Photonics (SPIE,
  2020), pp. 48 -- 55.

\bibitem{Knall_20_comp}
J.~Knall, M.~Engholm, J.~Ballato, P.~D. Dragic, N.~Yu, and M.~J.~F. Digonnet,
  \enquote{Experimental comparison of silica fibers for laser cooling,}
  {\protect\JournalTitle{Opt. Lett.}} \textbf{45}, 4020--4023 (2020).

\bibitem{zhou2016laser}
X.~Zhou, B.~E. Smith, P.~B. Roder, and P.~J. Pauzauskie, \enquote{Laser
  refrigeration of ytterbium-doped sodium--yttrium--fluoride nanowires,}
  {\protect\JournalTitle{Adv. Mater.}} \textbf{28}, 8658--8662 (2016).

\bibitem{zhu2010high}
X.~Zhu and N.~Peyghambarian, \enquote{High-power ZBALN glass fiber lasers:
  review and prospect,} {\protect\JournalTitle{Adv. Optoelectron.}}
  \textbf{2010} (2010).

\bibitem{mobini2019heatcc}
E.~Mobini, M.~Peysokhan, and A.~Mafi, \enquote{Heat mitigation of a core/cladding Yb-doped fiber amplifier using anti-Stokes fluorescence cooling,} {\protect\JournalTitle{J. Opt. Soc. Am. B}} \textbf{36}, 2167--2177 (2019).

\bibitem{jalali2006silicon}
B.~Jalali and S.~Fathpour, \enquote{Silicon photonics,} {\protect\JournalTitle{J. Lightwave Technol.}} \textbf{24}, 4600--4615 (2006).

\bibitem{soref2010mid}
R.~Soref, \enquote{Mid-infrared photonics in silicon and germanium,}
  {\protect\JournalTitle{Nat. Photonics}} \textbf{4}, 495--497 (2010).

\bibitem{sbowman}
S.~R. {Bowman}, \enquote{Lasers without internal heat generation,}
  {\protect\JournalTitle{IEEE J. Quant. Electron.}} \textbf{35},
  115--122 (1999).

\bibitem{NEMOVA20092571}
G.~Nemova and R.~Kashyap, \enquote{Athermal continuous-wave fiber amplifier,}
  {\protect\JournalTitle{Opt. Commun.}} \textbf{282}, 2571 -- 2575
  (2009).

\bibitem{bowman_min}
S.~R. {Bowman}, S.~P. {O'Connor}, S.~{Biswal}, N.~J. {Condon}, and
  A.~{Rosenberg}, \enquote{Minimizing heat generation in solid-state lasers,}
  {\protect\JournalTitle{IEEE J. Quant. Electron.}} \textbf{46},
  1076--1085 (2010).

\bibitem{Nemova:11}
G.~Nemova and R.~Kashyap, \enquote{Radiation-balanced amplifier with two pumps
  and a single system of ions,} {\protect\JournalTitle{J. Opt. Soc. Am. B}}
  \textbf{28}, 2191--2194 (2011).

\bibitem{Mobini:18}
E.~Mobini, M.~Peysokhan, B.~Abaie, and A.~Mafi, \enquote{Thermal modeling, heat
  mitigation, and radiative cooling for double-clad fiber amplifiers,}
  {\protect\JournalTitle{J. Opt. Soc. Am. B}} \textbf{35}, 2484--2493 (2018).

\bibitem{Yang:19}
Z.~Yang, J.~Meng, A.~R. Albrecht, and M.~Sheik-Bahae,
  \enquote{Radiation-balanced Yb:YAG disk laser,} {\protect\JournalTitle{Opt.
  Express}} \textbf{27}, 1392--1400 (2019).

\bibitem{Jesper}
J.~L{\ae}gsgaard, \enquote{Dissolution of rare-earth clusters in ${\rm SiO}_{2}$ by Al
  codoping: a microscopic model,} {\protect\JournalTitle{Phys. Rev. B}} \textbf{65}, 174114 (2002).

\bibitem{Arai}
K.~Arai, H.~Namikawa, K.~Kumata, T.~Honda, Y.~Ishii, and T.~Handa,
  \enquote{Aluminum or phosphorus co‐doping effects on the fluorescence and
  structural properties of neodymium‐doped silica glass,}
  {\protect\JournalTitle{J. Appl. Phys.}} \textbf{59}, 3430--3436
  (1986).

\bibitem{Rostami:19}
S.~Rostami, A.~R. Albrecht, A.~Volpi, and M.~Sheik-Bahae, \enquote{Observation
  of optical refrigeration in a holmium-doped crystal,}
  {\protect\JournalTitle{Photonics Res.}} \textbf{7}, 445--451 (2019).

\bibitem{yoder}
P.~Yoder, D.~Vukobratovich, and R.~A. Paquin, \emph{Opto-mechanical systems
  design, 2nd {E}d.} (CRC press, New York, 1992).

\bibitem{karimi2018theoretical}
M.~Karimi, \enquote{Theoretical study of the thermal distribution in {Yb}-doped
  double-clad fiber laser by considering different heat sources,}
  {\protect\JournalTitle{Prog. Electromagn. Res.}} \textbf{88},
  59--76 (2018).

\bibitem{mafi:20}
A.~Mafi, \enquote{Temperature distribution inside a double-cladding optical
  fiber laser or amplifier,} {\protect\JournalTitle{J. Opt. Soc. Am. B}}
  \textbf{37}, 1821--1828 (2020).

\bibitem{powerl1998physics}
R.~C.~Powerl, \enquote{Physics of solid-state laser materials,}  (Springer, New Yorl, 1998).

\bibitem{Peysokhan:19}
M.~Peysokhan, E.~Mobini, B.~Abaie, and A.~Mafi, \enquote{Method for measuring
  the resonant absorption coefficient of rare-earth-doped optical fibers,}
  {\protect\JournalTitle{Appl. Opt.}} \textbf{58}, 1841--1846 (2019).

\bibitem{Patterson:10}
W.~M. Patterson, D.~V. Seletskiy, M.~Sheik-Bahae, R.~I. Epstein, and M.~P.
  Hehlen, \enquote{Measurement of solid-state optical refrigeration by two-band
  differential luminescence thermometry,} {\protect\JournalTitle{J. Opt. Soc.
  Am. B}} \textbf{27}, 611--618 (2010).

\bibitem{Babakdlt}
B.~Imangholi, M.~P. Hasselbeck, D.~A. Bender, C.~Wang, M.~Sheik-Bahae, R.~I.
  Epstein, and S.~Kurtz, \enquote{{Differential luminescence thermometry in
  semiconductor laser cooling},} in \emph{Physics and Simulation of
  Optoelectronic Devices XIV,}  vol. 6115 M.~Osinski, F.~Henneberger, and
  Y.~Arakawa, eds., International Society for Optics and Photonics (SPIE,
  2006), pp. 215 -- 220.

\end{thebibliography}
\end{document}